\documentclass[a4paper, amsfonts, amssymb, amsmath, reprint, showkeys, nofootinbib, twoside,notitlepage,onecolumn]{revtex4-1}

\usepackage{amsmath,amstext}
\usepackage[T1]{fontenc}
\usepackage{amssymb}
\usepackage{graphicx}
\usepackage{ae,aecompl}

\DeclareFontFamily{OT1}{pzc}{}
\DeclareFontShape{OT1}{pzc}{m}{it}{<-> s * [1.10] pzcmi7t}{}
\DeclareMathAlphabet{\mathpzc}{OT1}{pzc}{m}{it}

\usepackage{hyperref}
\usepackage{amsmath}
\usepackage{amssymb}
\usepackage{mathtools}
\usepackage{bm}
\usepackage{cleveref}
\usepackage{tensor}
\usepackage{braket}
\usepackage{enumitem}
\usepackage{mhchem}
\usepackage{amsthm}
\usepackage{nccmath}
\usepackage{mathrsfs}
\usepackage{color}

\def\be{\begin{equation}}
\def\ee{\end{equation}}
\def\beq{\begin{eqnarray}}
\def\eeq{\end{eqnarray}}

\theoremstyle{definition}

\theoremstyle{theorem}

\theoremstyle{corollary}

\begin{document}
\title{Exact interpolation between Fick and Cattaneo diffusion in relativistic kinetic theory}
\author{L.~Gavassino}
\affiliation{Department of Applied Mathematics and Theoretical Physics, University of Cambridge, Wilberforce Road, Cambridge CB3 0WA, United Kingdom}

\begin{abstract}
We construct a family of exactly solvable relativistic kinetic theories in $1+1$ dimensions whose hydrodynamic sector continuously interpolates between Fick's and Cattaneo's laws of diffusion. The interpolation is controlled by a single parameter $a\in[0,1]$, which tunes the microscopic scattering dynamics from infinitely soft but infinitely frequent scatterings ($a=0$), reproducing standard diffusion, to maximally hard but finite-rate scatterings ($a=1$), yielding hyperbolic Cattaneo-type transport.
For intermediate values of $a$, the dynamics combines frequent weak scatterings with rare strong randomizing events, providing a concrete microscopic realization of mixed diffusive-telegraphic behavior. Remarkably, the full quasinormal mode spectrum can be obtained analytically for all $a$. This allows us to track explicitly how purely diffusive modes continuously deform into damped propagating modes as the collision structure is varied.
\end{abstract} 
\maketitle

\section{Introduction}

The standard view holds that Fick’s \cite{Fick1855} and Cattaneo’s \cite{cattaneo1958} laws of diffusion,
\begin{equation}\label{FickAndCattaneo}
\begin{split}
\partial_t n ={}& \mathfrak{D }\partial^2_x n \qquad\qquad\qquad \qquad (\text{Fick})\, ,\\
\partial_t n ={}&\mathfrak{D} (\partial^2_x-\partial^2_t) n \qquad\quad\qquad (\text{Cattaneo})\, ,\\
\end{split}
\end{equation}
are alternative macroscopic models for the same underlying transport physics \cite{Israel_Stewart_1979,MorseFeshbach1953,Jou_Extended,Muller_book,rezzolla_book,GavassinoAntonelli:2025umq}. According to this view, Fick’s law encodes the universal long-wavelength dynamics of a conserved density $n(t,x)$ propagating on a static background \cite[\S 9.3]{peliti_book}, in the regime where the associated current admits a first-order gradient expansion,
$J=-\mathfrak{D}\partial_x n+\mathcal{O}(\partial^2_x)$.
Cattaneo’s equation is then interpreted as a ``hyperbolic completion'' of Fick’s theory \cite{NagyOrtizReula1994,Geroch1995,GavassinoFronntiers2021}, incorporating a finite relaxation time for the current,
$\mathfrak{D}\partial_t J +J=-\mathfrak{D}\partial_x n$,
thereby enforcing causal propagation of density fluctuations.

Although this standard interpretation is certainly appropriate for generic physical systems, there exist scenarios in which the applicability of \eqref{FickAndCattaneo} can be pushed considerably further. In particular, one can identify microscopic models (see \cite{Amoretti2024,AhnBaggioli:2025odk}) where these equations are not merely \textit{effective} descriptions of the infrared dynamics, but instead govern the \textit{exact} evolution of the longest-lived excitation of the system, namely the hydrodynamic mode (i.e. the dispersion relation continuously connected to the origin in frequency space \cite{GavassinoNonHydro2022}). This is indeed the case for an ensemble of massless particles moving in one spatial dimension, randomly exchanging momentum with an external medium. In the limit where these exchanges are infinitely soft and infinitely frequent, the particle dynamics reduces to Fokker-Planck evolution in momentum space \cite{Debbasch,DunkelHanggi}. In this regime, one finds that the hydrodynamic mode obeys Fick’s law at all wavelengths \cite{GavassinoDiffusionCompatible2026tvy,GavassinoFokkerPlanck2026zsz}.
By contrast, when the mean free time between successive interactions with the medium is finite, but the collisions are fully randomizing, the kinetic description is captured by the Boltzmann equation in the Anderson-Witting relaxation-time approximation \cite{AndersonWitting1974}, which leads to an exact Cattaneo equation for the density \cite{Basar:2024qxd}. Hence, in this setting, the models \eqref{FickAndCattaneo} correspond to genuinely distinct physical behaviors, emerging as opposite limits of the microscopic collision structure, rather than as successive truncations of a single gradient expansion.

The existence of microscopic models reproducing \eqref{FickAndCattaneo} exactly is particularly valuable, as it enables one to address fundamental questions such as the causality and well-posedness of Fick’s law \cite{GavassinoDiffusionCompatible2026tvy,GavassinoLorentzBoostedDiffusion:2026fff}, or the consistent incorporation of stochastic fluctuations in Cattaneo’s theory. In the present work, we will focus on another intriguing aspect. 

The two equations in \eqref{FickAndCattaneo} are qualitatively different, one being parabolic and the other hyperbolic \cite{Kost2000}. The fact that both arise from closely related kinetic descriptions, as opposite limits of the scattering dynamics, naturally suggests the existence of an intermediate regime interpolating between these behaviors. More precisely, if a Fokker-Planck collision operator (associated with frequent soft scatterings) gives rise to Fick-type diffusion, while an Anderson-Witting collision term (associated with rare hard scatterings) leads to Cattaneo dynamics, then a weighted sum of the two should make it possible to construct a hydrodynamic mode influenced by both scattering mechanisms. Such a mode may then be continuously deformed from the parabolic Fick regime to the hyperbolic Cattaneo regime as the relative importance of soft versus hard scatterings is varied. As we shall see, the manner in which this mode deforms across regimes is highly nontrivial, and can be computed analytically.

Throughout the article, we work in natural units, with $c=\hbar=k_B=\text{Temperature of the environment}=1$.

\section{Setting the stage}
\vspace{-0.3cm}

In this preliminary section, we briefly review how the two equations \eqref{FickAndCattaneo} arise from non-degenerate ultrarelativistic kinetic theories in $1+1$ dimensions. We will follow the derivations given in \cite{GavassinoDiffusionCompatible2026tvy,GavassinoFokkerPlanck2026zsz,Basar:2024qxd}. 

\vspace{-0.3cm}
\subsection{The Boltzmann equation}
\vspace{-0.3cm}

We consider a dilute gas of massless particles constrained to move along a line. If $p$ denotes the linear momentum of a particle, its energy is $\varepsilon(p)=|p|$, and its velocity is $v(p)=d\varepsilon/dp=\text{sign}(p)$. The particles do not interact among themselves, but undergo inelastic scattering with an external medium\footnote{The external medium selects a preferred inertial frame and therefore breaks formal Lorentz covariance of the theory. Accordingly, our models are ``relativistic'' in the sense that the particles obey relativistic kinematics and information propagates exactly at the speed of light. However, Lorentz boosts are not a symmetry of the collision operator.} held at temperature $1$ (in our units). The phase-space distribution function $f(t,x,p)$ then satisfies the Boltzmann equation \cite{cercignani_book,Groot1980RelativisticKT}
\begin{equation}\label{Boltzmann}
(\partial_t + v \partial_x) f = \mathcal{C}\, ,
\end{equation}
where $\mathcal{C}[f]$ denotes the collision term\footnote{It is worth noting that, in a strictly $(1+1)$--dimensional setting, closed systems of colliding particles are integrable. As a consequence, thermalization does not occur, the molecular chaos assumption fails, and the collision term $\mathcal{C}$ cannot be expressed as a functional of the one-particle distribution $f$ alone, since higher-order correlations must be retained. This would pose a difficulty if the scattering processes originated from binary interactions among particles within the gas itself. In the present setup, however, we are dealing with an open kinetic system, so momentum exchange is mediated by an external medium, which may be assumed to carry additional degrees of freedom that break integrability and allow for ergodic behavior. For instance, one may envision a population of particles (or quasi-particles) constrained to a one-dimensional channel and interacting with an external three-dimensional thermal reservoir. All in all, the analysis only requires the existence of a Markovian environment giving rise to the effective collision operator $\mathcal{C}[f]$.}.
Because the particles do not interact with one another and the bath is assumed to be Markovian, the collision term $\mathcal{C}$ is linear in $f$ \cite{GavassinoMemoryEffects:2026ebx}. This linearity plays a crucial role in what follows, as it implies that the solutions of \eqref{Boltzmann} can be decomposed exactly into linear superpositions of excitation modes.

The collision term has the following properties:
\begin{equation}\label{conditionsOnC}
\begin{split}
\int_\mathbb{R} \dfrac{dp}{2\pi} \mathcal{C}=0 & \qquad \qquad (\text{particle conservation})\, , \\
\mathcal{C}[e^{-\varepsilon}]=0 & \qquad \qquad(\text{existence of equilibrium})\, . \\
\end{split}
\end{equation}
The first condition guarantees that, if we integrate both sides of \eqref{Boltzmann} over all momenta, we obtain the continuity equation $\partial_t n+\partial_x J=0$, where the particle density and the particle flux are given by \cite[\S I.1.a]{Groot1980RelativisticKT}
\begin{equation}
\begin{bmatrix}
n \\
J \\
\end{bmatrix}
=
\int_{\mathbb{R}} \dfrac{dp}{2\pi} \begin{bmatrix}
1 \\
v \\
\end{bmatrix}
f \, .
\end{equation}
The second condition in \eqref{conditionsOnC} guarantees that, whenever the particles are in thermal equilibrium with the environment (i.e., have its same temperature), the scatterings do not have any net effect on the phase space distribution.

\vspace{-0.3cm}
\subsection{The Fick limit}
\vspace{-0.3cm}

If the momentum exchanges between a particle and the environment are infinitely soft but infinitely frequent, the particle undergoes Brownian motion in momentum space. Such a process necessarily includes a drift term, ensuring that the second condition in \eqref{conditionsOnC} is satisfied. The corresponding collision operator takes the Fokker-Planck form \cite{Debbasch,DunkelHanggi}
\begin{equation}\label{FokkerPLanck}
\mathcal{C}[f]=\nu \partial_p (\partial_p f + v f)\, ,
\end{equation}
where $\nu=\text{const}>0$ is the momentum diffusivity. The first condition in \eqref{conditionsOnC} is also fulfilled, provided that $f$ decays sufficiently rapidly at large $p$, ensuring finiteness of the density.

To establish that the hydrodynamic mode exhibits Fickian diffusion, we observe that the ansatz
\begin{equation}\label{ansatz}
f=e^{-\varepsilon+ik(x-p/\nu)-i\omega t}
\end{equation}
solves the Boltzmann equation \eqref{Boltzmann} with collision operator \eqref{FokkerPLanck}, provided $\omega=-ik^2/\nu$. Consequently, linear superpositions of modes of the form \eqref{ansatz} generate densities $n(t,x)$ that solve exactly the first equation in \eqref{FickAndCattaneo}, with $\mathfrak{D}=\nu^{-1}$, assuming convergence of the $k$ and $p$ integrals.

An alternative derivation follows from
\begin{equation}
0=-\int_{\mathbb{R}}\frac{dp}{2\pi}\,\partial_p f
=\int_{\mathbb{R}}\frac{dp}{2\pi}(v+ik/\nu)f
=J+\frac{ik}{\nu}n
=J+\nu^{-1}\partial_x n\, ,
\end{equation}
i.e., Fick's constitutive relation $J=-\mathfrak{D}\partial_x n$ is recovered.
Here we again require $f$ to decay at large $p$, which implies $|\mathfrak{Im}\,k|<\nu$. Violation of this condition leads to divergent density and hence to unphysical modes\footnote{Although Fickian diffusion, when regarded as a standalone partial differential equation, is acausal and unstable, there is no obstruction to its arising as the exact hydrodynamic sector of a causal and stable microscopic theory, as shown in \cite{GavassinoDiffusionCompatible2026tvy,GavassinoHowacausal:2026fil}. For $k\notin\mathbb{R}$, however, this embedding must be understood only conditionally, since the unstable modes that appear in boosted frames cannot be reproduced by a causal and stable microscopic theory. In the case of Fokker-Planck, the necessary and sufficient condition for the embedding to exist is precisely the convergence of the momentum integral $\int_\mathbb{R} \frac{dp}{2\pi} f$, which automatically removes all unstable modes \cite{GavassinoLorentzBoostedDiffusion:2026fff}.}.

\subsection{The Cattaneo limit}\label{CattaneoForEver}

We now turn to the opposite regime, in which particles propagate ballistically over a mean free time $\Gamma^{-1}=\text{const}>0$, followed by fully randomizing scattering events. In this case, the collision term takes the Anderson-Witting form \cite{AndersonWitting1974}
\begin{equation}\label{andersonwitting}
\mathcal{C}[f]=\Gamma (f_\text{eq}-f)\, .
\end{equation}
The equilibrium distribution $f_\text{eq}(t,x,p)$ is fixed by imposing both conditions in \eqref{conditionsOnC}: it must possess the same particle density of $f$ and be locally thermal with the bath, implying proportionality to $e^{-\varepsilon}$. This yields
\begin{equation}\label{fequil}
f_\text{eq}=\pi n e^{-\varepsilon}\, .
\end{equation}

To extract the density dynamics, we introduce the densities of right- and left-moving particles \cite{Basar:2024qxd},
$n_+{=}\int_0^{+\infty}\frac{dp}{2\pi}f$ and
$n_-{=}\int_{-\infty}^0\frac{dp}{2\pi}f$,
which propagate with velocities $+1$ and $-1$, respectively. The total density and flux are $n=n_++n_-$ and $J=n_+-n_-$. Integrating both sides of \eqref{Boltzmann} over positive and negative momenta and using \eqref{andersonwitting}–\eqref{fequil}, we obtain
\begin{equation}
\begin{split}
(\partial_t+\partial_x)n_+ ={}& -\frac{\Gamma}{2}J\, ,\\
(\partial_t-\partial_x)n_- ={}& +\frac{\Gamma}{2}J\, .
\end{split}
\end{equation}
Summing these equations yields the continuity equation $\partial_t n+\partial_x J=0$, while subtracting them gives Cattaneo’s relaxation law $\Gamma^{-1}\partial_t J+J=-\Gamma^{-1}\partial_x n$. Combining the two reproduces the second line of \eqref{FickAndCattaneo}, with $\mathfrak{D}=\Gamma^{-1}$.

This derivation does not rely on any assumption regarding the structure of $f$, beyond finiteness of the density. This means that \textit{every} solution of Cattaneo’s equation admits a kinetic embedding, and \textit{every} kinetic solution obeys Cattaneo's law. In particular, besides the hydrodynamic mode, Cattaneo’s theory also contains a non-hydrodynamic branch, which captures exactly the transient relaxation of the density predicted by the underlying kinetic dynamics\footnote{The two Cattaneo modes do not exhaust the full kinetic spectrum: additional non-hydrodynamic modes with $n=J=0$ are present, to which the Cattaneo equation is insensitive.}.

\begin{figure}[h!]
    \centering
\includegraphics[width=0.45\linewidth]{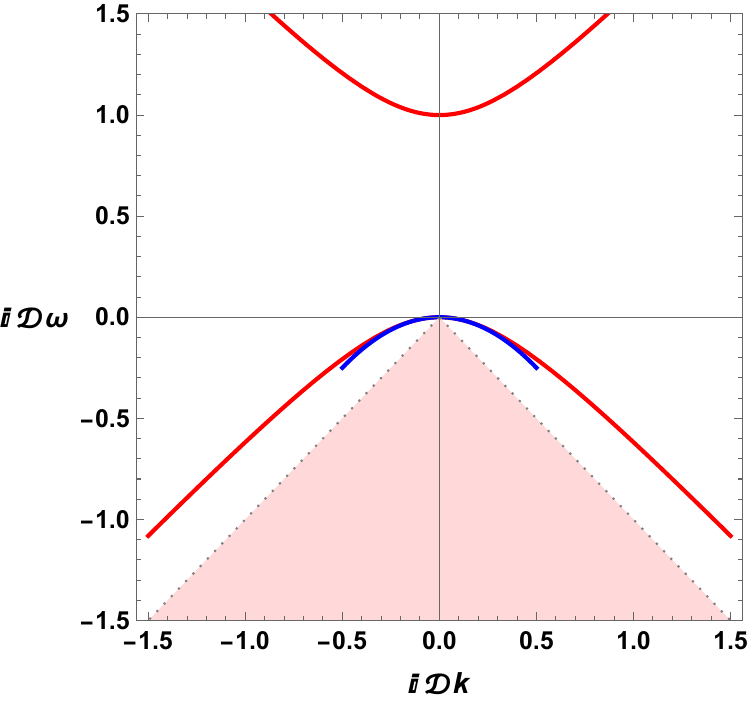}\hspace{0.08\linewidth}
\includegraphics[width=0.45\linewidth]{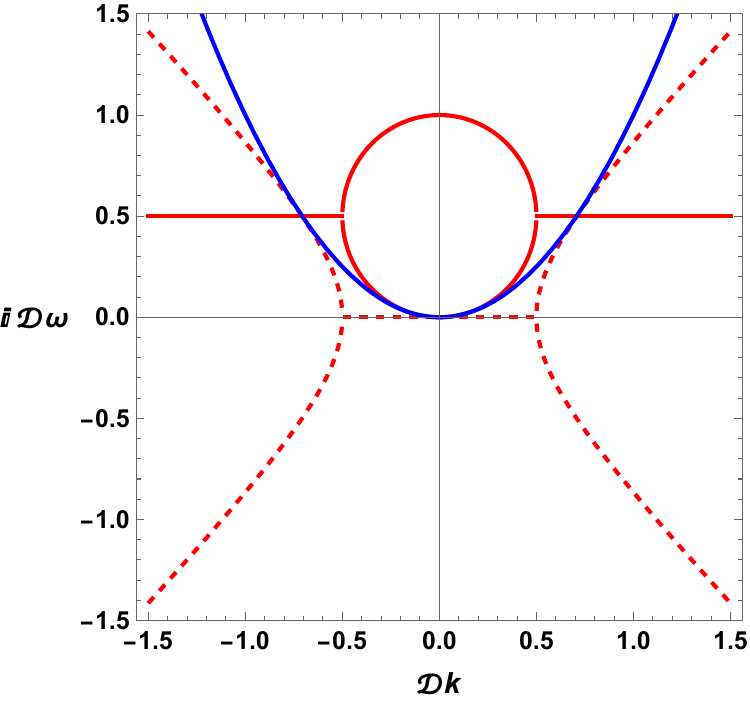}
\caption{Dispersion relations $\omega=-i\mathfrak{D}k^2$ and $\omega=-i\mathfrak{D}(k^2-\omega^2)$ for Fick (blue) and Cattaneo (red) diffusion, respectively, as obtained from the corresponding kinetic theories. Left panel: Frequency plotted as a function of imaginary $k$. The Fick branch terminates at $|\mathfrak{Im}k|=(2\mathfrak{D})^{-1}$, beyond which the information current associated with the mode \eqref{ansatz} diverges, rendering the mode unphysical (the particle density itself diverges at $|\mathfrak{Im}k|=\mathfrak{D}^{-1}$). By contrast, the Cattaneo branches extend to arbitrarily large $|k|$. As expected, none of the curves enter the region $\mathfrak{Im}\omega<|\mathfrak{Im}k|$ (shaded), where the modes would become unstable under Lorentz boosts \cite{GavassinoBounds2023myj}. Right panel: Frequency plotted as a function of real $k$. Both Fick and Cattaneo branches extend over the full real axis. In the Cattaneo case, $i\mathfrak{D}\omega$ develops an imaginary part (red dashed) for $|k|>(2\mathfrak{D})^{-1}$, indicating the onset of propagating behavior \cite{Pu2010,BAGGIOLI20201,GavassinoGENERIC:2022isg}. In both cases, one finds $\mathfrak{Im}\omega\leq 0$, consistent with linear stability.
}
    \label{fig:DispFromPDEs}
\end{figure}

In figure \ref{fig:DispFromPDEs}, we graph the dispersion relations of the modes arising from \eqref{FickAndCattaneo}, keeping track of their regime of applicability when embedded within the corresponding kinetic theories.

\section{The interpolating theory}

We are finally entering the central part of the paper, where we explicitly compute and analyze the modes of a kinetic theory whose collision integral is the sum of a Fokker-Planck term \eqref{FokkerPLanck} and an Anderson-Witting term \eqref{andersonwitting}, with appropriate weights.

\subsection{Schr\"odinger representation}

We start from the Boltzmann equation
\begin{equation}\label{interpuls}
(\partial_t + v \partial_x) f = \nu \partial_p (\partial_p f + v f)-\Gamma f+\Gamma e^{-\varepsilon} \int_{\mathbb{R}} \dfrac{d\Tilde{p}}{2}
\Tilde{f}\, ,
\end{equation}
where $\Tilde{f}$ denotes $f$ evaluated at momentum $\Tilde{p}$. Upon adopting the ansatz
$f(t,x,p)=e^{-\varepsilon/2+ikx-i\omega t}\psi(p)$, which is standard in problems involving a Fokker-Planck operator \cite[\S 5.4, \S 5.5]{RiskenFP},
we obtain
\begin{equation}\label{themonster}
-\nu \psi''+[ik\, \text{sign}(p)-\nu\delta(p) ]\psi-\Gamma e^{-\varepsilon/2} \int_{\mathbb{R}} \dfrac{d\Tilde{p}}{2} e^{-\Tilde{\varepsilon}/2}\Tilde{\psi} =\left(i\omega-\dfrac{\nu}{4}-\Gamma \right)\psi \, .
\end{equation}
This equation may be viewed as a one-dimensional Schr\"odinger problem, with $p$ playing the role of the spatial coordinate\footnote{The natural Hilbert space of this Schr\"odinger problem, namely $L^2(\mathbb{R})$, is the space of functions with a finite information current \cite{DudynskiEkielJezewska1985,GavassinoCausality2021,GavassinoGapless:2024rck,RochaGavassinoFlucut:2024afv}, and the inner product
$\langle\phi|\psi\rangle=\int_{\mathbb{R}}\phi^*\psi\,dp$
is the Hessian of the grand potential, and thus coincides with Onsager's inner product \cite{GavassinoDistrubingMoving:2026klp,GavassinoFokkerPlanck2026zsz}. In this sense, $L^2(\mathbb{R})$ provides the natural space of functions to work with.}, and with an effective potential given by the sum of a $\text{sign}(p)$ term, a $\delta(p)$ interaction, and a rank-one projector $\ket{e^{-\varepsilon/2}}\bra{e^{-\varepsilon/2}}$. When $k$ is purely imaginary, the resulting Hamiltonian is self-adjoint, and the eigenvalue $i\omega-\nu/4-\Gamma$ is real. For general $k$, however, the Hamiltonian becomes non-Hermitian, and the spectrum is generically complex. In this representation, equilibrium corresponds to $\psi\propto e^{-\varepsilon/2}$, with $\omega=k=0$.

\subsection{The continuous spectrum suggests a natural parameterization}

Our main goal is to determine the hydrodynamic mode for this family of theories, which corresponds to a discrete eigenvalue (i.e. an element of the point spectrum) of the effective Schr\"odinger Hamiltonian. By contrast, the continuous part of the spectrum is comparatively straightforward to characterize. For operators of the present type (with $\nu>0$), the continuous spectrum coincides with the essential spectrum, which may be extracted by considering wavefunctions supported asymptotically at $p\to\pm\infty$, and solving the limiting Schr\"odinger equation in that regime \cite[\S 6.4]{TeschlBook}. In our case, this yields
$-\nu \psi''=(i\omega-\nu/4-\Gamma\pm ik)\psi$,
and therefore the continuous branches $i\omega \in \pm ik +\big[\frac{\nu}{4}+\Gamma,\infty \big)$.

This observation motivates the parametrization
\begin{equation}\label{parameterization}
\nu =\dfrac{4 (1-a)}{\tau}  \, , \qquad \qquad \Gamma=\dfrac{a}{\tau} \, .
\end{equation}
With this choice of parameters, if we keep $\tau$ fixed and vary $a$ over the interval $[0,1]$, the continuous spectrum remains fixed (except at $a=1$, where it collapses to $i\omega=\pm ik+1/\tau$), and takes the universal form
\begin{equation}\label{continuousspectrumaless1}
i\omega \in \pm ik +\bigg[\dfrac{1}{\tau},\infty \bigg)\, ,
\end{equation}
while the relative weight of the Fokker-Planck and Anderson-Witting terms is tuned continuously from pure Fokker-Planck dynamics at $a=0$ to pure Anderson-Witting relaxation at $a=1$. In this language, $\tau$ plays the role of the microscopic relaxation time, whereas $a$ quantifies the relative importance of the strongly randomizing scattering events. Note that, when $a=1$, $\tau$ coincides with the standard Anderson-Witting relaxation time $\Gamma^{-1}$.

Plugging \eqref{parameterization} into \eqref{themonster}, introducing the notation $i\omega\tau=E$ and $\chi=ik\tau$, and fixing the normalization of $\psi$ so that $\int_{\mathbb{R}} \frac{d\Tilde{p}}{2} e^{-\Tilde{\varepsilon}/2}\Tilde{\psi}=1$, we arrive at the central equation of this work:
\begin{equation}\label{Lequazionefondamentale}
\boxed{
-4 (1{-}a) \psi''+[1-E+\chi\, \text{sign}(p)-4 (1{-}a)\delta(p)]\psi =a e^{-\varepsilon/2}  \, .}
\end{equation}

\newpage
\subsection{Bound-state solutions}

To solve \eqref{Lequazionefondamentale}, we treat separately the regions $p<0$ and $p>0$, obtaining
\begin{equation}\label{brokendown}
\begin{split}
-4 (1{-}a) \psi''+(1-E-\chi)\psi ={}& a e^{+p/2} \qquad\qquad (p<0) \, , \\
-4 (1{-}a) \psi''+(1-E+\chi)\psi ={}& a e^{-p/2} \qquad\qquad (p>0) \, . \\
\end{split}
\end{equation}
These equations are supplemented by the matching conditions at $p=0$,
\begin{equation}\label{matching}
\psi(0^+)=\psi(0^-)\, , \qquad \qquad \psi'(0^+)-\psi'(0^-)+\psi(0)=0 \, ,
\end{equation}
which follow from integrating across the $\delta(p)$ interaction.

To determine the hydrodynamic mode (which, we recall, is a discrete eigenvalue), we search for bound states, i.e. solutions $\psi(p)$ that decay as $p\to\pm\infty$ (rather than merely oscillating). Since the equations \eqref{brokendown} are linear and inhomogeneous, the solution in each half-line is the sum of a particular solution proportional to the source term $e^{\pm p/2}$ and of the general solution of the associated homogeneous equation. The latter is a (generically complex) exponential. Imposing decay at infinity, we may parameterize the homogeneous pieces as $e^{\lambda_1 p}$ for $p<0$ and $e^{-\lambda_2 p}$ for $p>0$, with $\mathfrak{Re}\lambda_{1,2}>0$.
These exponents are related to $E$ and $\chi$ by substituting the homogeneous ansatz into \eqref{brokendown}, which gives
\begin{equation}\label{EandChi}
\begin{split}
E={}& -2(1-a) \left(\lambda_1^2+\lambda_2^2 \right)+1 \, ,\\
\chi={}& -2(1-a) \left(\lambda_1^2-\lambda_2^2 \right) \, .\\
\end{split}
\end{equation}
Accordingly, we take
\begin{equation}
\psi =
\begin{cases}
A_1 e^{+p/2}+B_1 e^{+\lambda_1 p} & p<0 \, ,\\
A_2 e^{-p/2}+B_2 e^{-\lambda_2 p} & p>0 \, ,\\
\end{cases}
\end{equation}
and impose \eqref{brokendown} together with the matching conditions \eqref{matching}. This yields
\begin{equation}\label{A1A2B1B2}
A_1= \dfrac{a}{a{-}E{-}\chi} \,, \quad \, \, \, \,
B_1= -\dfrac{a\chi}{(a{-}E)^2-\chi^2} \, \dfrac{1{-}2\lambda_2}{1{-}\lambda_1{-}\lambda_2} \,,\quad \, \, \, \,
A_2= \dfrac{a}{a{-}E{+}\chi} \,, \quad \, \, \, \,
B_2=\dfrac{a\chi}{(a{-}E)^2-\chi^2} \, \dfrac{1{-}2\lambda_1}{1{-}\lambda_1{-}\lambda_2} \,.
\end{equation}
There is one last constraint, which comes from the fact that, to arrive at \eqref{Lequazionefondamentale}, we have assumed that
$\int_{\mathbb{R}} \frac{dp}{2} e^{-|p|/2}\psi=1$. This requirement translates into the following identity:
\begin{equation}
\dfrac{A_1+A_2}{2} +\dfrac{B_1}{1+2\lambda_1}+\dfrac{B_2}{1+2\lambda_2}-1=0 \, .
\end{equation}
Using \eqref{EandChi} and \eqref{A1A2B1B2}, the left-hand side becomes a rational function of $\lambda_1$, $\lambda_2$, and $a$. Clearing denominators leads to a polynomial constraint of the form $\mathcal P(\lambda_1,\lambda_2,a)=0$, where $\mathcal P$, regarded as a polynomial in either $\lambda_1$ or $\lambda_2$, is of degree three. Although this equation admits an explicit analytic solution, both $\mathcal P$ and its roots are rather cumbersome, and we therefore refrain from displaying them here.

\subsection{Imaginary wavenumber}

We recall that, when $\chi=ik\tau$ is real, the effective Hamiltonian is self-adjoint, implying that $E=i\omega\tau$ is likewise real. Combined with the bound-state condition $\mathfrak{Re}\,\lambda_{1,2}>0$, this shows that modes with imaginary wavenumber are characterized by real, positive $\lambda_{1,2}$.

\begin{figure}[h!]
    \centering
\includegraphics[width=0.45\linewidth]{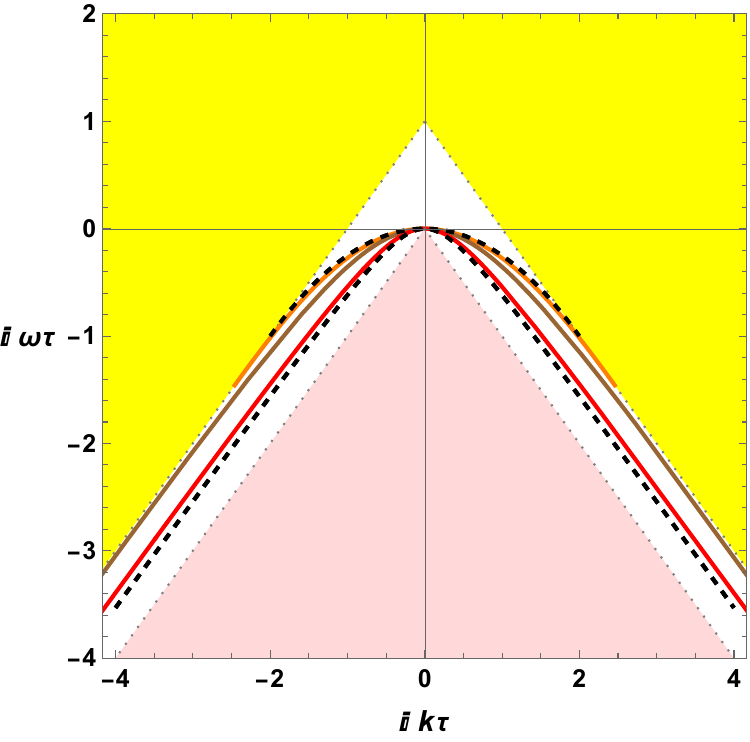}\hspace{0.05\linewidth}
\includegraphics[width=0.46\linewidth]{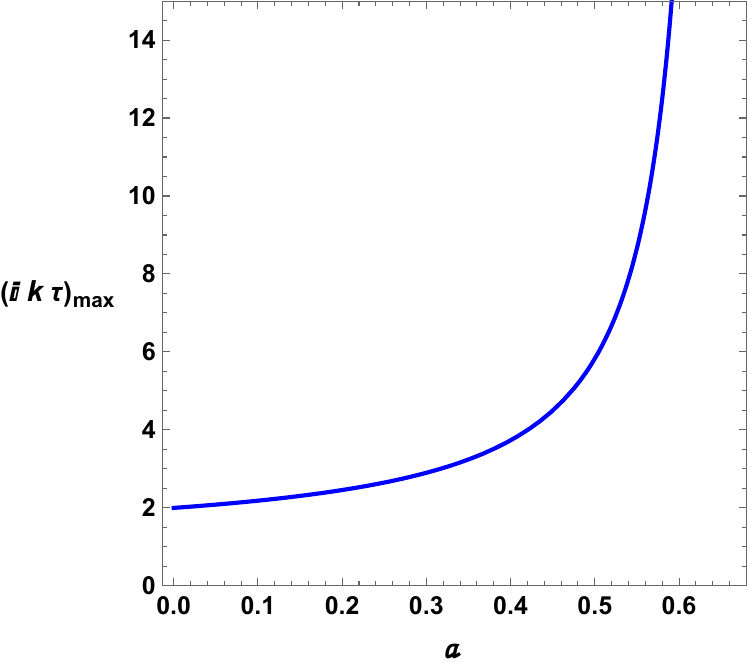}
\caption{Hydrodynamic dispersion relation $\omega(k)$ of the Fokker-Planck-Anderson-Witting kinetic theory \eqref{interpuls} for imaginary wave number and various values of the parameter $a$ (see equation \eqref{parameterization}). 
Left panel: As $a$ increases (orange: $a=1/5$, brown: $a=2/3$, red: $a=0.98$), the dispersion relation interpolates continuously between Fick’s law (upper dashed curve), with $\mathfrak{D}=\tau/4$, and Cattaneo’s law (lower dashed curve), with $\mathfrak{D}=\tau$. Throughout this deformation, the continuous spectrum remains fixed and occupies the region $i\omega\geq -|ik|+1/\tau$ (yellow shading). Consistently with causality \cite{HellerBounds2022ejw,GavassinoDistrubingMoving:2026klp}, none of the branches enters the unstable region $\mathfrak{Im}\omega<|\mathfrak{Im}k|$ (red shading).
Right panel: For $a<2/3$, the hydrodynamic branch merges with the continuous spectrum at a finite value $(ik\tau)_{\text{max}}$, beyond which it ceases to exist. In particular, at $a=0$ one recovers Fick’s law, which is defined only for $|\mathfrak{Im}(k\tau)|<2$. As $a$ increases, $(ik\tau)_{\text{max}}$ grows monotonically and diverges as $a\to 2/3$, signaling the transition to the Cattaneo regime, where the dispersion relation extends over the entire imaginary axis.
}
    \label{fig:ImaginaryInterpolation}
\end{figure}

With this in mind, we construct the dispersion relation at imaginary wavenumber as follows. We solve the constraint $\mathcal{P}(\lambda_1,\lambda_2,a)=0$ for $\lambda_2$, obtaining three branches. Letting $\lambda_1$ vary over $[0,\infty)$, we find that only one branch yields $\lambda_2>0$. Each admissible pair $(\lambda_1,\lambda_2)$ is then substituted into equation \eqref{EandChi}, providing a parametric representation of the dispersion relation. The resulting curves are shown in figure \ref{fig:ImaginaryInterpolation} (left panel).

Interestingly, for $a<2/3$, the function $\lambda_2(\lambda_1)$ remains positive only over a finite interval of $\lambda_1$, and is bounded. Consequently, the resulting dispersion relation $\omega(k)$ is defined only over a finite range of imaginary wavenumbers, where the maximal value of $ik\tau$ (which coincides with the point where the mode is tangent to the continuous spectrum) is plotted in figure \ref{fig:ImaginaryInterpolation} (right panel). By contrast, for $a\geq 2/3$, the function $\lambda_2(\lambda_1)$ is positive on a half-line $(\lambda_1^{\text{(min)}},\infty)$ and diverges near $\lambda_1^{\text{(min)}}$, which causes the resulting parametric curve $(ik(\lambda_1),i\omega(\lambda_1))$ to explore all values of $k\in i\mathbb{R}$. This behavior is consistent with the transition from Fick’s law, which applies only for $|\mathfrak{Im}k|\leq(2\mathfrak{D})^{-1}$, to Cattaneo’s law, which remains valid for arbitrary imaginary $k$ (see section \ref{CattaneoForEver}).

\subsection{The diffusivity coefficient}
\vspace{-0.2cm}

Figure~\ref{fig:ImaginaryInterpolation} shows that the diffusivity $\mathfrak{D}$ (defined as the coefficient in the small-$k$ expansion $\omega=-i\mathfrak{D}k^2+\dots$) increases monotonically with $a$. In particular, at $a=0$ one recovers Fick’s law with $\mathfrak{D}=\tau/4$, while at $a=1$ one obtains Cattaneo’s law with $\mathfrak{D}=\tau$. Remarkably, the interpolating diffusivity $\mathfrak{D}(a)$ admits a simple closed-form expression.

To determine $\mathfrak{D}(a)$, we note that by definition
\vspace{-0.1cm}
\begin{equation}
\dfrac{\mathfrak{D}}{\tau}=-\frac{1}{2}\frac{d^2E}{d\chi^2}\bigg|_{\chi=0}\, .
\end{equation}
The condition $\chi=0$ (for which $E=0$) is attained for $\lambda_1=\lambda_2=(2\sqrt{1-a})^{-1}$, as follows directly from equation \eqref{EandChi}. Treating $E$ and $\chi$ as functions of $\lambda_1$, one may write
\vspace{-0.1cm}
\begin{equation}
\dfrac{\mathfrak{D}}{\tau}=\frac{\chi''E'-\chi'E''}{2(\chi')^3}\, ,
\end{equation}
where primes denote \textit{total} derivatives with respect to $\lambda_1$, evaluated at $\lambda_1=(2\sqrt{1-a})^{-1}$. These derivatives are
\vspace{-0.1cm}
\begin{equation}\label{EandChiDeriv}
\begin{split}
E'={}& -4(1-a)\lambda_1\left(1+\lambda_2' \right)\, ,\\
E''={}& -4(1-a)\left[1+(\lambda_2')^2+\lambda_2\lambda_2'' \right]\, ,\\
\chi'={}& -4(1-a)\lambda_1\left(1-\lambda_2' \right)\, ,\\
\chi''={}& -4(1-a)\left[1-(\lambda_2')^2-\lambda_2\lambda_2'' \right]\, .
\end{split}
\end{equation}
Exploiting the symmetry of $\mathcal{P}(\lambda_1,\lambda_2,a)$ under $\lambda_1\leftrightarrow\lambda_2$, we infer that $\lambda_2'=-1$ at $\lambda_1=\lambda_2$, which implies $E'=0$ and $\chi'=-4\sqrt{1-a}$. This reduces the expression for the diffusivity to
$\mathfrak{D}=\frac{\tau}{8}\bigl(2+\lambda_2\lambda_2''\bigr)$.
The remaining second derivative $\lambda_2''$ is obtained by applying the implicit function theorem to $\mathcal{P}(\lambda_1,\lambda_2,a)=0$ about $\lambda_1=\lambda_2=(2\sqrt{1-a})^{-1}$, yielding
\vspace{-0.1cm}
\begin{equation}
\mathfrak{D}(a)=\frac{\tau}{2-a+2\sqrt{1-a}}\, .
\end{equation}
As expected, this reproduces $\mathfrak{D}(0)=\tau/4$ and $\mathfrak{D}(1)=\tau$.

\subsection{Non-propagating modes at real wavenumber}
\vspace{-0.2cm}

For most physical applications, it is natural to focus on modes with real wavenumber $k$, since these provide the Fourier basis from which localized wavepackets may be constructed (at least in the rest frame \cite{Hiscock_Insatibility_first_order,GavassinoSuperlum2021}). As shown in figure~\ref{fig:DispFromPDEs} (right panel), the non-propagating excitations, characterized by $i\omega\in\mathbb{R}$, arrange themselves along a parabola in the Fick limit and along a circle in the Cattaneo limit. We now examine how the kinetic model~\eqref{interpuls} continuously interpolates between these two geometries.

\begin{figure}[h!]
    \centering
\includegraphics[width=0.46\linewidth]{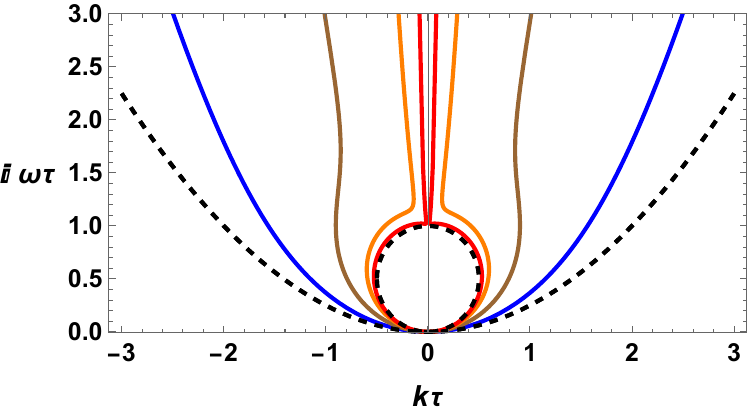}\hspace{0.05\linewidth}
\includegraphics[width=0.45\linewidth]{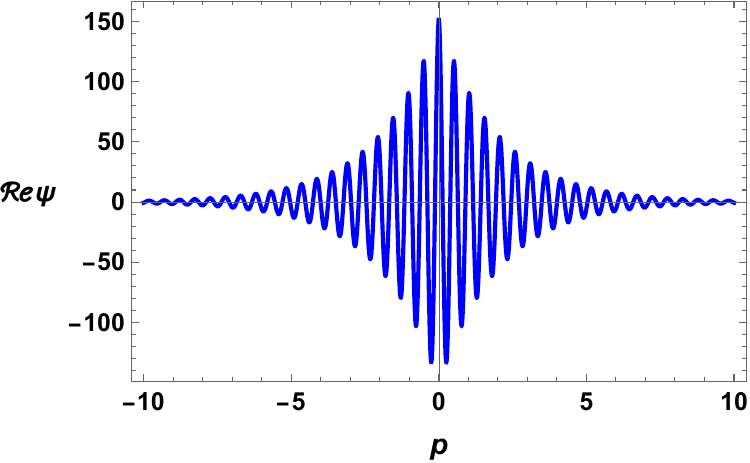}
\caption{Left panel: Non-propagating branch of the discrete dispersion relation $\omega(k)$ of the interpolating kinetic theory~\eqref{interpuls} for real wavenumber and various values of the parameter $a$ (blue: $a=0.5$, brown: $a=0.9$, orange: $a=0.99$, red: $a=0.999$). As $a$ increases, the curve interpolates continuously between the parabolic Fick geometry and the circular Cattaneo geometry. The non-propagating modes, characterized by $i\omega\in\mathbb{R}$, are obtained by setting $\lambda_1=r-is$ and $\lambda_2=r+is$ with $r>0$ and exploring the corresponding parametric curve $(k(r),\omega(r))$ over the interval $1/2<r\le (2\sqrt{1-a})^{-1}$. For all $a<1$, $\chi$ spans the entire imaginary axis, and the curves extend to arbitrarily large $|k|$.
In the limit $a\to 1$, the branch approaches a circle of radius $1$, together with a vertical half-line emerging from $(k\tau,i\omega\tau)=(0,1)$. This additional line is absent in the Cattaneo theory ($a\equiv 1$) and reflects an exchange-of-limits issue. If one fixes $r\in[1/2+\varepsilon,(2\sqrt{1-a})^{-1}]$ with $\varepsilon>0$ and then sends $a\to1$, only the Cattaneo circle survives. If instead $a$ is fixed and $r\to1/2$, then $s$ diverges and the wavefunction $\psi(p)$ becomes highly oscillatory (see right panel for an explicit example). In this ultraviolet regime, the diffusive Fokker-Planck term~\eqref{FokkerPLanck} dominates, even when the momentum diffusivity $\nu$ is small, causing the mode to decay on a timescale much shorter than $\tau$. For the same reason, the continuous spectrum is given by~\eqref{continuousspectrumaless1} for all $a<1$, while it collapses to $\pm ik+1/\tau$ at $a=1$: arbitrarily oscillatory states are always damped by the Fokker-Planck term at finite $1-a$. Right panel: Real part of the effective wavefunction $\psi(p)$ corresponding to $a=0.9$ and $r=1/2+0.0001$, illustrating the highly oscillatory ultraviolet behavior near the endpoint $r=1/2$.}
    \label{fig:NonPropagating}
\end{figure}

We decompose the parameters $\lambda_{1,2}$ into their real and imaginary parts: $\lambda_1=r_1+is_1$ and $\lambda_2=r_2+is_2$, with $r_{1,2}>0$ to ensure boundedness. Requiring $\chi$ to be imaginary and $E$ to be real, equation~\eqref{EandChi} implies the constraints $r_1^2-r_2^2=s_1^2-s_2^2$ and $r_1s_1+r_2 s_2=0$. The only real solutions compatible with $r_{1,2}>0$ are
\begin{equation}
r_1=r_2\equiv r \, , \qquad \qquad s_1=-s_2\equiv -s \, .
\end{equation}
Accordingly, one must determine the real roots $s(r)$ of the polynomial $\mathcal{P}(r{+}is,r{-}is,a)$, which reduces to the quartic form $c_0+c_2 s^2+c_4 s^4$, with
\begin{equation}
\begin{split}
c_0 ={}& -(2 r+1)^3 \left[4 (a-1) r^2+1\right] \, ,\\
c_2 ={}& -4 (2 r-1) \left[8 a r (r+1)+a-2 (2 r+1)^2\right] \, , \\
c_4 ={}& -16 (a-1) (2 r-1) \, . \\
\end{split}
\end{equation}
Among the four roots, there are only two real ones (with opposite signs), restricted to the interval $1/2 {<} r{\leq} (2\sqrt{1{-}a})^{-1}$, within which they interpolate from $\pm\infty$ to zero. Consequently, $\chi=-8i(1-a)rs$ spans the entire imaginary axis, while $E=4(1-a)(s^2-r^2)+1$ explores all non-negative values. The resulting deformation of the non-propagating branch is displayed in figure~\ref{fig:NonPropagating} (left panel).

As $a$ increases, Fick’s parabola $i\omega=\tau k^2/4$ progressively shrinks and deforms around Cattaneo’s circle $i\omega\,{=}\,\tau(k^2{-}\omega^2)$. Notably, for all $a<1$ there remains a large-$\omega$ non-propagating branch with a parabolic-like profile, extending over the entire real axis $k\in\mathbb{R}$. This branch is non-hydrodynamic. In the limit $a\to1$, it collapses into a vertical half-line at $k=0$, corresponding to modes that become arbitrarily oscillatory in momentum space (see figure~\ref{fig:NonPropagating}, right panel), for which the Fokker-Planck term dominates even when $\nu$ is arbitrarily small. Overall, while the infrared geometry of the hydrodynamic branch is continuously deformed as $a$ varies, the ultraviolet structure of the spectrum remains Fokker-Planck-like for all $a<1$.

\subsection{Propagating modes at real wavenumber}

In the previous section, we analyzed the branch of real-wavenumber modes that are non-propagating (i.e.\ $i\omega\in\mathbb{R}$). However, Cattaneo’s theory is known to admit, in addition, a branch of propagating modes, while Fick’s theory does not (see figure~\ref{fig:DispFromPDEs}, right panel). We now examine how this feature emerges in the interpolating kinetic theory.

\begin{figure}[h!]
    \centering
\includegraphics[width=0.35\linewidth]{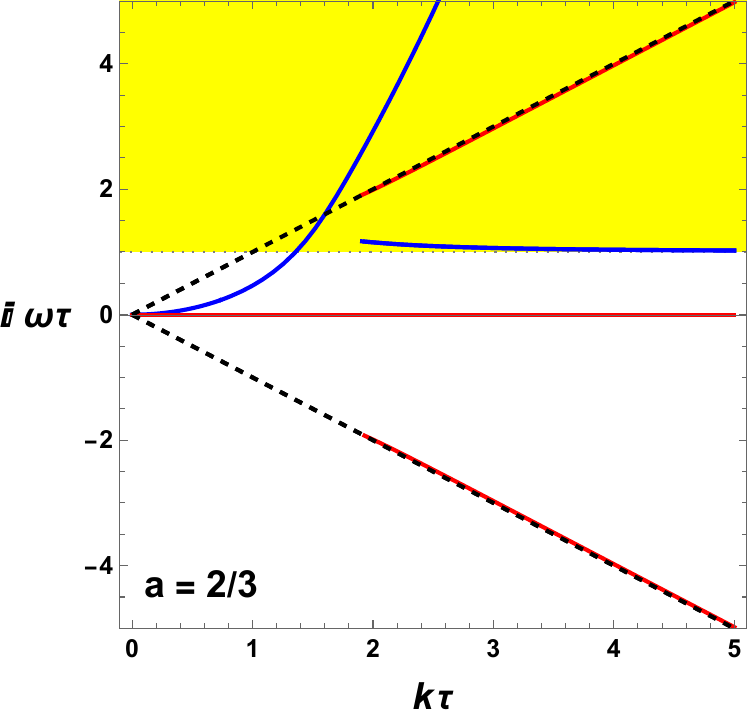}\hspace{0.05\linewidth}
\includegraphics[width=0.35\linewidth]{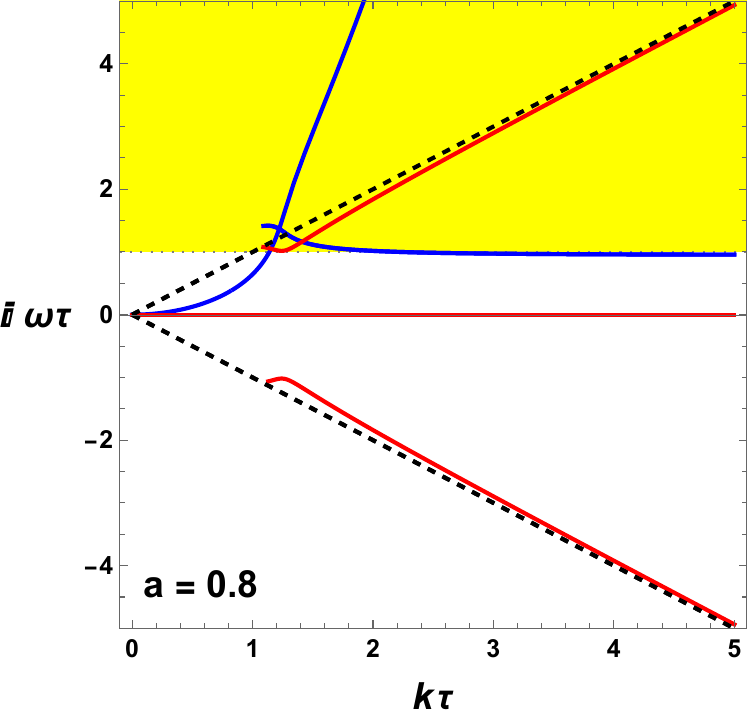}
\includegraphics[width=0.35\linewidth]{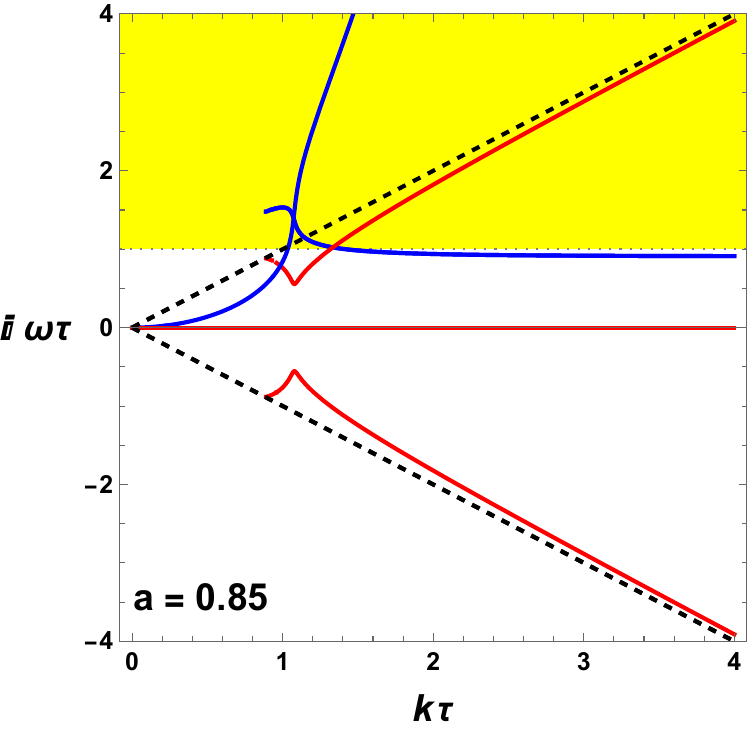}\hspace{0.05\linewidth}
\includegraphics[width=0.35\linewidth]{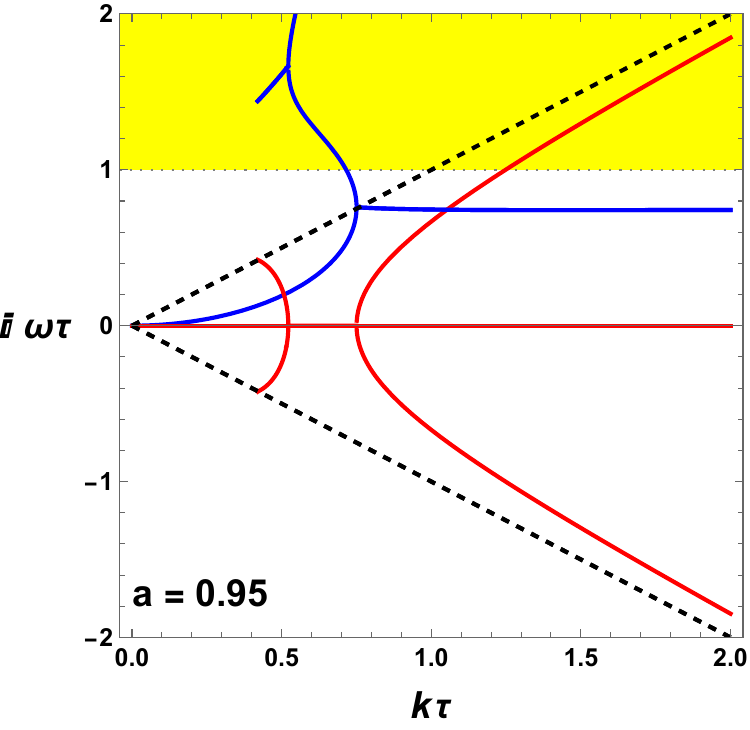}
\caption{Real (blue) and imaginary (red) parts of the dispersion relations $i\omega(k)$ for real $k$, including both the non-propagating branch (cf.\ figure~\ref{fig:NonPropagating}) and the propagating branches, shown for representative values of $a$. A pair of propagating modes bifurcates from the continuous spectrum, where the latter occupies the region $\mathfrak{Re}\,(i\omega)\ge 1/\tau$ (yellow~shading) and $\mathfrak{Im}\,(i\omega)=\pm k$ (black~dashed), at a finite value of $k$. For $a<a_{\text{critical}}\approx 0.87$, this pair persists up to $k\to\infty$. For $a\ge a_{\text{critical}}$, the propagating modes are absorbed by the non-propagating branch at the forward tilt point and subsequently reemerge at the backward tilt point (note that $a_{\text{critical}}$ is the value of $a$ at which the non-propagating dispersion relation ceases to be single-valued). In the Cattaneo limit, only the branch emanating from the backward tilt point remains.}
    \label{fig:Propagating}
\end{figure}

Unlike in the preceding case, it is no longer advantageous to explicitly use the fact that $\chi$ is imaginary, since $E$ is generically complex and decomposing $\lambda_{1,2}$ into real and imaginary parts only obscures the root structure. Instead, to enforce the reality of $k$, we adopt a different strategy. From the second equation of \eqref{EandChi}, we obtain\footnote{Here, the square root is taken on the principal branch (as in \textsc{Mathematica}), for which $\mathfrak{Re}\lambda_2\ge 0$. For $\mathfrak{Re}\lambda_1>0$ and $\chi\in i\mathbb{R}$, this choice selects the physically admissible solution, while the opposite branch has $\mathfrak{Re}\lambda_2\le 0$ and is therefore unphysical.}
\begin{equation}
\lambda_2=\sqrt{\lambda_1^2+\dfrac{\chi}{2(1-a)}} \qquad \qquad (\chi \in i\mathbb{R}) \, .
\end{equation}
Substituting this relation into $\mathcal{P}(\lambda_1,\lambda_2,a)$ and rearranging the resulting expression, we find that all solutions are also roots of a polynomial $\mathcal{Q}(\lambda_1,\chi,a)$ of degree $9$ in $\lambda_1$. Many of these roots are unphysical (either they do not satisfy $\mathcal{P}=0$, or they have $\mathfrak{Re}\lambda_1<0$), and which branches are admissible depends on the value of $\chi$. The procedure is therefore straightforward (albeit tedious): we plot $E(\chi)=1-\chi-4(1-a)\lambda_1(\chi)^2$ for all nine branches, impose the physicality conditions, and discard the unacceptable ones. The resulting dispersion relations are shown in figure~\ref{fig:Propagating}.

Numerically, we find evidence for the existence of a lower threshold $a_\ast=0.6221\pm0.0001$,
below which no discrete propagating modes are present. More precisely, for $a<a_\ast$, the only bound-state solution corresponds to a purely damped excitation, whereas a conjugate pair with $\mathfrak{Re}\,\omega\neq0$ appears for $a>a_\ast$. Close to the threshold, the propagating branch exists only over a finite interval of real wavenumbers centered around $k\tau\approx4.7$, which broadens continuously as $a$ increases. Although no closed-form expression for $a_\ast$ is presently known (being the solution of an inverse problem), its numerical value can be determined to arbitrary precision.

\vspace{-0.3cm}
\section{Conclusions}
\vspace{-0.3cm}

A natural interpolation between Fick and Cattaneo diffusion at the level of partial differential equations is provided by the parametric Cattaneo model \begin{equation}\label{cattaneowitha} \partial_t n=\mathfrak{D}(\partial_x^2-a\,\partial_t^2)n, \end{equation} which continuously deforms the spectrum by pushing the non-hydrodynamic mode towards infinity as $a\searrow0$. Although this interpolation is mathematically straightforward, its microscopic realization is less obvious. Indeed, realizing it kinetically would appear to require simultaneously increasing the particle propagation speed and the non-hydrodynamic gap frequency. We are not aware of a kinetic construction that reproduces equation \eqref{cattaneowitha} by merely changing the scattering mechanism while keeping the particle propagation (sub)luminal and the maximal relaxation timescale fixed.

The construction developed in the present work follows a fundamentally different philosophy. Throughout the interpolation, particles always propagate at the speed of light, while the non-hydrodynamic gap remains fixed. The only ingredient that changes is the microscopic scattering mechanism, namely the relative importance of infinitely frequent soft momentum transfers and rare complete randomization events. The resulting one-parameter family therefore provides a genuine microscopic interpolation between Fick and Cattaneo diffusion, making it possible to isolate the consequences of changing the collision dynamics while leaving the kinematics unchanged.

The spectrum undergoes a correspondingly rich deformation. Along the imaginary $k$-axis, the hydrodynamic branch continuously evolves from Fick's parabola to Cattaneo's hyperbola, covering the entire imaginary axis for $a\geq2/3$ (figure \ref{fig:ImaginaryInterpolation}). Along the real $k$-axis, the non-propagating branch deforms continuously from the Fick parabola to the Cattaneo circle (figure \ref{fig:NonPropagating}), while retaining, for every $a<1$, an ultraviolet branch inherited from the Fokker--Planck dynamics. This branch disappears only in the strict Anderson--Witting limit, revealing a non-trivial exchange of limits between short-wavelength oscillations of the momentum distribution and vanishing soft-scattering rate. Meanwhile, a pair of propagating modes appears once the interpolation parameter exceeds a threshold $a_\ast\approx 0.6221$, and for $a\gtrsim0.87$ is absorbed and subsequently reemitted by the non-propagating branch at its vertical inflection points (figure \ref{fig:Propagating}).

These results provide a microscopic explanation for the qualitative difference between Fick and Cattaneo diffusion. The transition from Fick's instantaneous smoothing (characteristic of parabolic transport) to Cattaneo's damped propagation (characteristic of hyperbolic transport) is not driven by a change in the microscopic causal structure, since every kinetic theory in the interpolating family is causal (see Appendix \ref{aaa} for the proof). Instead, it is controlled by the nature of the scattering mechanism. Infinitely frequent soft momentum transfers continuously damp short-wavelength perturbations, generating a hierarchy of rapidly decaying diffusive modes that disappears only in the strict Anderson--Witting limit. As these collisions are progressively replaced by rare complete randomization events, this smoothing mechanism is gradually lost, allowing propagating behavior to emerge. Frequent soft collisions are therefore substantially more effective at smoothing disturbances (both in physical space and in momentum space) than rare but maximally randomizing collisions, even when the maximal relaxation timescale is kept fixed.

Although the present analysis was restricted to one spatial dimension, we expect the overall physical picture to persist in $3+1$ dimensions. Both the relativistic Fokker--Planck and Anderson--Witting collision operators admit natural higher-dimensional generalizations, and the mapping of the spectral problem onto an equivalent Schr\"odinger equation remains available. The main technical difference is that the effective potential generated by the Fokker--Planck operator becomes the three-dimensional Coulomb potential \cite{GavassinoFokkerPlanck2026zsz}, replacing the discontinuous one-dimensional barrier studied here. On the hydrodynamic side, the Anderson--Witting model no longer reproduces the Cattaneo equation exactly, but rather a non-local constitutive relation. Nevertheless, its qualitative transport properties remain the same, suggesting that the microscopic interpolation introduced here should continue to provide a useful framework for understanding the transition between Fick-like and Cattaneo-like diffusion in more realistic settings.

\section*{Acknowledgements}

This work is supported by a MERAC Foundation prize grant,  an Isaac Newton Trust Grant, and funding from the Cambridge Centre for Theoretical Cosmology.

\appendix

\section{Proof of causality of the interpolating kinetic theories}\label{aaa}

In this appendix, we prove that the kinetic equation \eqref{interpuls} is causal in the sense of \cite{DudynskiCausality}. In particular, if $f(0,x,p)=0$ for $|x|<L$, then $f(t,x,p)=0$ for $|x|<L-t$ for any $t>0$.

Following \cite{GavassinoCausality2021}, we introduce the positive current
\begin{equation}
E^\mu =\dfrac{1}{2}\int_\mathbb{R} \dfrac{dp}{2\pi}
\begin{bmatrix}
1 \\
v \\
\end{bmatrix}
e^{\varepsilon}f^2 \, .
\end{equation}
This vector is finite whenever $e^{\varepsilon/2}f\sim\psi$ is square integrable. Moreover, since $|v|\le1$, it is manifestly future-directed and non-spacelike. Multiplying equation \eqref{interpuls} by $e^\varepsilon f$, integrating over all momenta, and integrating by parts in $p$, one finds $\partial_\mu E^\mu=-\sigma$, where
\begin{equation}
\sigma = \nu \int \dfrac{dp}{2\pi} e^\varepsilon (\partial_p f+ v f)^2 +\Gamma \int \dfrac{dp}{2\pi} e^\varepsilon \left( f-e^{-\varepsilon}\int \dfrac{d\Tilde{p}}{2}\Tilde{f} \right)^2 \geq 0 \, .
\end{equation}
The non-negativity of $\sigma$ follows immediately from the fact that it is the sum of squares.

\begin{figure}[h!]
    \centering
\includegraphics[width=0.42\linewidth]{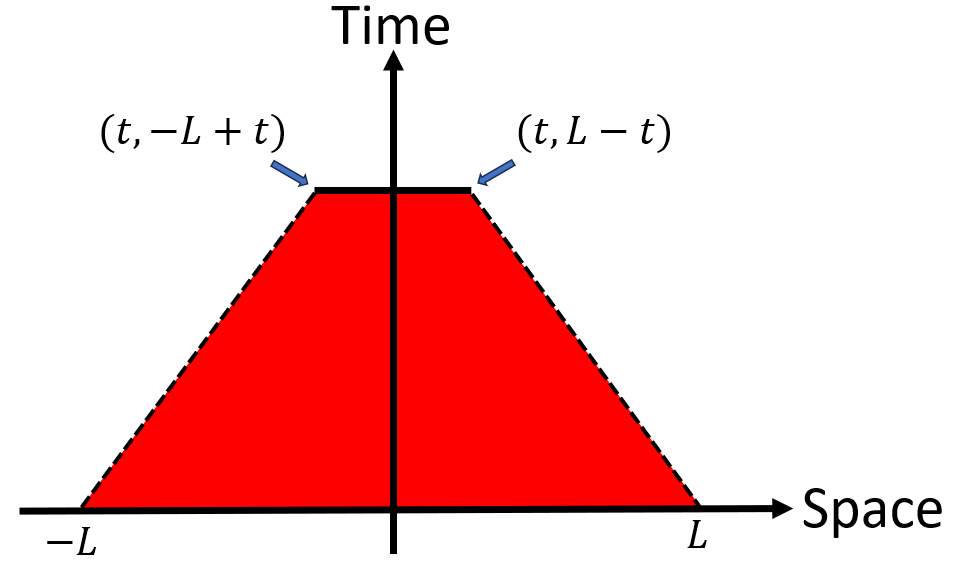}
\caption{Integration domain used in the proof of causality. The balance law $\partial_\mu E^\mu+\sigma=0$ is integrated over the trapezium bounded by the initial interval $[-L,L]$, the two null characteristics $x=\pm L\mp t$, and the final interval $[-L+t,L-t]$. Since the flux through the null boundaries and the entropy production $\sigma$ are both non-negative, vanishing initial data imply vanishing data on the upper base, establishing the domain-of-dependence property.}
    \label{fig:CausalityProof}
\end{figure}

We now integrate the identity $\partial_\mu E^\mu+\sigma=0$ over the trapezium shown in Fig.~\ref{fig:CausalityProof}. Applying Gauss' theorem yields
\begin{equation}
\begin{split}
&\int_{\text{trapezium}} \sigma(t',x')\, dt' dx'+ \int_{-L+t}^{L-t} E^0(t,x)\, dx-\int_{-L}^{L} E^0(0,x)\, dx \\
&+\int_{-L}^{-L+t} [E^0(L{+}x,x)-E^1(L{+}x,x)]\,dx
+\int_{L-t}^{L} [E^0(L{-}x,x)+E^1(L{-}x,x)]\,dx=0 \, .
\end{split}
\end{equation}
Since $E^\mu$ is future-directed and non-spacelike, we have $E^0\pm E^1\geq0$, so the outward flux through the slanted sides of the trapezium is non-negative. Likewise, $\sigma\ge0$, so the volume contribution is also non-negative. Finally, because $f$ vanishes on the lower base of the trapezium by assumption, we have $E^0(0,x)=0$ for $x\in[-L,L]$. Therefore,
\begin{equation}
\int_{-L+t}^{L-t} E^0(t,x)\, dx\leq 0.
\end{equation}
Since $E^0\ge0$ by construction, the above inequality implies that $E^0(t,x)=0$ almost everywhere on the upper base of the trapezium. By definition,
\begin{equation}
E^0=\frac12\int_\mathbb{R}\frac{dp}{2\pi}e^\varepsilon f^2,
\end{equation}
which vanishes only if $f=0$. Hence, $f(t,x,p)=0$ for $|x|<L-t$, as claimed. 

The above result, combined with the linearity of the kinetic equation, immediately implies the full domain-of-dependence property \cite[\S 10.1]{Wald}. Indeed, let $f_1(t,x,p)$ and $f_2(t,x,p)$ be two solutions with identical initial data for $|x|<L$. Then their difference $f_1-f_2$ is itself a solution, and vanishes initially on that interval. By the result just proved, it must therefore vanish for $|x|<L-t$. Hence, the two solutions coincide throughout this spacetime region. It follows that no disturbance can propagate outside the light cone, so the interpolating kinetic theories are causal.

\bibliography{Biblio}

@article{GavassinoLorentzBoostedDiffusion:2026fff,
    author = "Gavassino, Lorenzo",
    title = "{Lorentz-boosted diffusion: Initial value formulation and exact solutions}",
    eprint = "2602.21254",
    archivePrefix = "arXiv",
    primaryClass = "math-ph",
    doi = "10.1103/k646-mgz5",
    journal = "Phys. Rev. D",
    volume = "113",
    number = "11",
    pages = "114038",
    year = "2026"
}

@article{GavassinoDiffusionCompatible2026tvy,
    author = "Gavassino, Lorenzo",
    title = "{Diffusion Equation is Compatible with Special Relativity}",
    eprint = "2601.19464",
    archivePrefix = "arXiv",
    primaryClass = "gr-qc",
    doi = "10.1103/rkz9-xps2",
    journal = "Phys. Rev. Lett.",
    volume = "137",
    number = "2",
    pages = "022302",
    year = "2026"
}

@article{GavassinoFokkerPlanck2026zsz,
    author = "Gavassino, Lorenzo",
    title = "{Quasinormal modes of relativistic Fokker-Planck kinetic theory}",
    eprint = "2601.19474",
    archivePrefix = "arXiv",
    primaryClass = "nucl-th",
    doi = "10.1103/6vfx-8kmm",
    journal = "Phys. Rev. D",
    volume = "114",
    number = "1",
    pages = "014018",
    year = "2026"
}

@book{RiskenFP,
  author    = {H. Risken},
  title     = {The Fokker--Planck Equation},
  publisher = {Springer},
  year      = {1989}
}

@article{Debbasch,
  author  = {Debbasch, F. and Mallick, K. and Rivet, J.-P.},
  title   = {Relativistic Ornstein--Uhlenbeck process},
  journal = {Journal of Statistical Physics},
  volume  = {88},
  number  = {4},
  pages   = {945--966},
  year    = {1997},
  doi     = {10.1023/B:JOSS.0000015180.16261.53},
}

@article{DunkelHanggi,
    author = {Dunkel, J{\"o}rn and H{\"a}nggi, Peter},
    title = "{Relativistic Brownian motion}",
    eprint = "0812.1996",
    archivePrefix = "arXiv",
    primaryClass = "cond-mat.stat-mech",
    doi = "10.1016/j.physrep.2008.12.001",
    journal = "Phys. Rept.",
    volume = "471",
    pages = "1--73",
    year = "2009"
}

@article{GavassinoDistrubingMoving:2026klp,
    author = "Gavassino, Lorenzo",
    title = "{How Lorentz Boosts Reshape Relaxation Spectra}",
    eprint = "2601.03081",
    archivePrefix = "arXiv",
    primaryClass = "gr-qc",
    doi = "10.1103/d2xr-rj7g",
    journal = "Phys. Rev. Lett.",
    volume = "137",
    number = "2",
    pages = "022301",
    year = "2026"
}

@article{RochaGavassinoFlucut:2024afv,
    author = "Soares Rocha, Gabriel and Gavassino, Lorenzo and Mullins, Nicki",
    title = "{Modeling stochastic fluctuations in relativistic kinetic theory}",
    eprint = "2405.10878",
    archivePrefix = "arXiv",
    primaryClass = "nucl-th",
    doi = "10.1103/PhysRevD.110.016020",
    journal = "Phys. Rev. D",
    volume = "110",
    number = "1",
    pages = "016020",
    year = "2024"
}

@article{DudynskiEkielJezewska1985,
  author  = {Dudy{\'n}ski, Marek and Ekiel-Jezewska, Maria L.},
  title   = {On the Linearized Relativistic Boltzmann Equation. I. Existence of Solutions},
  journal = {Communications in Mathematical Physics},
  volume  = {102},
  pages   = {17--34},
  year    = {1985}
}

@article{GavassinoBounds2023myj,
    author = "Gavassino, L.",
    title = "{Bounds on transport from hydrodynamic stability}",
    eprint = "2301.06651",
    archivePrefix = "arXiv",
    primaryClass = "hep-th",
    doi = "10.1016/j.physletb.2023.137854",
    journal = "Phys. Lett. B",
    volume = "840",
    pages = "137854",
    year = "2023"
}

@article{Basar:2024qxd,
    author = {Ba\c{s}ar, G\"ok\c{c}e and Bhambure, Jay and Singh, Rajeev and Teaney, Derek},
    title = "{Stochastic relativistic advection diffusion equation~from the Metropolis algorithm}",
    eprint = "2403.04185",
    archivePrefix = "arXiv",
    primaryClass = "nucl-th",
    doi = "10.1103/PhysRevC.110.044903",
    journal = "Phys. Rev. C",
    volume = "110",
    number = "4",
    pages = "044903",
    year = "2024"
}

@ARTICLE{GavassinoNonHydro2022,
     author = "Gavassino, Lorenzo and Antonelli, Marco and Haskell, Brynmor",
    title = "{Symmetric-hyperbolic quasihydrodynamics}",
    eprint = "2207.14778",
    archivePrefix = "arXiv",
    primaryClass = "gr-qc",
    doi = "10.1103/PhysRevD.106.056010",
    journal = "Phys. Rev. D",
    volume = "106",
    number = "5",
    pages = "056010",
    year = "2022"
}

@ARTICLE{Pu2010,
       author = {{Pu}, Shi and {Koide}, Tomoi and {Rischke}, Dirk H.},
        title = "{Does stability of relativistic dissipative fluid dynamics imply causality?}",
      journal = {\prd},
         year = 2010,
        month = jun,
       volume = {81},
       number = {11},
          eid = {114039},
        pages = {114039},
          doi = {10.1103/PhysRevD.81.114039},
archivePrefix = {arXiv},
       eprint = {0907.3906},
 primaryClass = {hep-ph},
       adsurl = {https://ui.adsabs.harvard.edu/abs/2010PhRvD..81k4039P}
}

@ARTICLE{GavassinoSuperlum2021,
    author = "Gavassino, Lorenzo",
    title = "{Can We Make Sense of Dissipation without Causality?}",
    eprint = "2111.05254",
    archivePrefix = "arXiv",
    primaryClass = "gr-qc",
    doi = "10.1103/PhysRevX.12.041001",
    journal = "Phys. Rev. X",
    volume = "12",
    number = "4",
    pages = "041001",
    year = "2022"
}

@ARTICLE{Geroch1995,
       author = {{Geroch}, Robert},
        title = "{Relativistic theories of dissipative fluids}",
      journal = {Journal of Mathematical Physics},
         year = 1995,
        month = aug,
       volume = {36},
       number = {8},
        pages = {4226-4241},
          doi = {10.1063/1.530958},
       adsurl = {https://ui.adsabs.harvard.edu/abs/1995JMP....36.4226G}
}

@article{GavassinoGENERIC:2022isg,
    author = "Gavassino, Lorenzo and Antonelli, Marco",
    title = "{Relativistic liquids: GENERIC or EIT?}",
    eprint = "2209.12865",
    archivePrefix = "arXiv",
    primaryClass = "gr-qc",
    doi = "10.1088/1361-6382/acc165",
    journal = "Class. Quant. Grav.",
    volume = "40",
    number = "7",
    pages = "075012",
    year = "2023"
}

@ARTICLE{GavassinoFronntiers2021,
    author = "Gavassino, Lorenzo and Antonelli, Marco",
    title = "{Unified Extended Irreversible Thermodynamics and the stability of relativistic theories for dissipation}",
    eprint = "2105.15184",
    archivePrefix = "arXiv",
    primaryClass = "gr-qc",
    doi = "10.3389/fspas.2021.686344",
    journal = "Front. Astron. Space Sci.",
    volume = "8",
    pages = "686344",
    year = "2021"
}

@ARTICLE{GavassinoCausality2021,
    author = "Gavassino, Lorenzo and Antonelli, Marco and Haskell, Brynmor",
    title = "{Thermodynamic Stability Implies Causality}",
    eprint = "2105.14621",
    archivePrefix = "arXiv",
    primaryClass = "gr-qc",
    doi = "10.1103/PhysRevLett.128.010606",
    journal = "Phys. Rev. Lett.",
    volume = "128",
    number = "1",
    pages = "010606",
    year = "2022"
}

@book{Groot1980RelativisticKT,
  title={Relativistic kinetic theory: principles and applications},
  author={S. R. de Groot and W. Astrid van Leeuwen and Christianus G. van Weert},
  year={1980}
}

@book{cattaneo1958,
  title={Sur une forme de l'{\'e}quation de la chaleur {\'e}liminant le paradoxe d'une propagation instantan{\'e}e},
  author={Cattaneo, C. },
  series={Comptes rendus hebdomadaires des s{\'e}ances de l'Acad{\'e}mie des sciences},
  url={https://books.google.pl/books?id=mHGeQwAACAAJ},
  year={1958},
  publisher={Gauthier-Villars}
}

@ARTICLE{Kost2000,
       author = {{Kost{\"a}dt}, Peter and {Liu}, Mario},
        title = "{Causality and stability of the relativistic diffusion equation}",
      journal = {\prd},
         year = 2000,
        month = jul,
       volume = {62},
       number = {2},
          eid = {023003},
        pages = {023003},
          doi = {10.1103/PhysRevD.62.023003},
archivePrefix = {arXiv},
       eprint = {cond-mat/0010276},
 primaryClass = {cond-mat.stat-mech},
       adsurl = {https://ui.adsabs.harvard.edu/abs/2000PhRvD..62b3003K}
}

@BOOK{cercignani_book,
       author = {{Cercignani}, Carlo and {Kremer}, Gilberto Medeiros},
        title = "{The relativistic Boltzmann equation: theory and applications}",
         year = "2002",
       adsurl = {https://ui.adsabs.harvard.edu/abs/2002rbet.book.....C}
}

@article{GavassinoHowacausal:2026fil,
    author = "Gavassino, Lorenzo",
    title = "{How acausal equations emerge from causal dynamics}",
    eprint = "2604.07031",
    archivePrefix = "arXiv",
    primaryClass = "nucl-th",
    month = "4",
    year = "2026"
}

@article{GavassinoMemoryEffects:2026ebx,
    author = "Gavassino, Lorenzo and Mitra, Sukanya and Singh, Rajeev",
    title = "{Hydrodynamics without a relaxation gap: Memory effects, nonlocality, and superdiffusion}",
    eprint = "2606.01805",
    archivePrefix = "arXiv",
    primaryClass = "nucl-th",
    doi = "10.1103/9362-8gmq",
    journal = "Phys. Rev. D",
    volume = "114",
    number = "3",
    pages = "034038",
    year = "2026"
}

@article{DudynskiCausality,
  title = {Causality of the Linearized Relativistic Boltzmann Equation},
  author = {Dudynski, Marek and Ekiel-Jezewska, Maria L.},
  journal = {Phys. Rev. Lett.},
  volume = {56},
  issue = {20},
  pages = {2228},
  numpages = {0},
  year = {1986},
  month = {May},
  publisher = {American Physical Society},
  doi = {10.1103/PhysRevLett.56.2228},
  url = {https://link.aps.org/doi/10.1103/PhysRevLett.56.2228}
}

@book{Wald,
      author        = "Wald, Robert M",
      title         = "{General relativity}",
      publisher     = "Chicago Univ. Press",
      address       = "Chicago, IL",
      year          = "1984",
      url           = "https://cds.cern.ch/record/106274",
}

@article{Hiscock_Insatibility_first_order,
author = {Hiscock, William and Lindblom, Lee},
year = {1985},
month = {03},
pages = {725-733},
title = {Generic instabilities in first-order dissipative relativistic fluid theories},
volume = {31},
journal = {Physical review D: Particles and fields},
doi = {10.1103/PhysRevD.31.725}
}

@article{GavassinoGapless:2024rck,
    author = "Gavassino, Lorenzo",
    title = "{Gapless nonhydrodynamic modes in relativistic kinetic theory}",
    eprint = "2404.12327",
    archivePrefix = "arXiv",
    primaryClass = "nucl-th",
    doi = "10.1103/PhysRevResearch.6.L042043",
    journal = "Phys. Rev. Res.",
    volume = "6",
    number = "4",
    pages = "L042043",
    year = "2024"
}

@article{Amoretti2024,
  title = {Relaxed hydrodynamic theory of electrically driven nonequilibrium steady states},
  author = {Brattan, Daniel K. and Matsumoto, Masataka and Baggioli, Matteo and Amoretti, Andrea},
  journal = {Phys. Rev. Res.},
  volume = {6},
  issue = {4},
  pages = {043097},
  numpages = {14},
  year = {2024},
  month = {Nov},
  publisher = {American Physical Society},
  doi = {10.1103/PhysRevResearch.6.043097},
  url = {https://link.aps.org/doi/10.1103/PhysRevResearch.6.043097}
}

@article{AhnBaggioli:2025odk,
    author = "Ahn, Yongjun and Baggioli, Matteo and Bu, Yanyan and Matsumoto, Masataka and Sun, Xiyang",
    title = "{Simple holographic dual of the Maxwell-Cattaneo model and the fate of KMS symmetry for nonhydrodynamic modes}",
    eprint = "2506.00926",
    archivePrefix = "arXiv",
    primaryClass = "hep-th",
    doi = "10.1103/3kx1-156x",
    journal = "Phys. Rev. D",
    volume = "112",
    number = "8",
    pages = "086013",
    year = "2025"
}

@article{Israel_Stewart_1979,
title = "Transient relativistic thermodynamics and kinetic theory",
journal = "Annals of Physics",
volume = "118",
number = "2",
pages = "341 - 372",
year = "1979",
issn = "0003-4916",
doi = "https://doi.org/10.1016/0003-4916(79)90130-1",
url = "http://www.sciencedirect.com/science/article/pii/0003491679901301",
author = "W. Israel and J.M. Stewart"
}

@article{GavassinoAntonelli:2025umq,
     author = "Gavassino, Lorenzo and Antonelli, Marco",
    title = "{Heat propagation in rotating relativistic bodies}",
    eprint = "2509.00198",
    archivePrefix = "arXiv",
    primaryClass = "gr-qc",
    doi = "10.1103/4g4w-ypmf",
    journal = "Phys. Rev. D",
    volume = "112",
    number = "10",
    pages = "104052",
    year = "2025"
}

@article{Jou_Extended,
author = {Jou, David and Casas-V\'{a}zquez, Jos\'{e} and Lebon, Georgy},
year = {1999},
month = {01},
pages = {1105},
title = {Extended Irreversible Thermodynamics},
volume = {51},
journal = {Reports on Progress in Physics},
doi = {10.1088/0034-4885/51/8/002}
}

@ARTICLE{AndersonWitting1974,
       author = {{Anderson}, J.~L. and {Witting}, H.~R.},
        title = "{A relativistic relaxation-time model for the Boltzmann equation}",
      journal = {Physica},
         year = 1974,
        month = jun,
       volume = {74},
       number = {3},
        pages = {466-488},
          doi = {10.1016/0031-8914(74)90355-3},
       adsurl = {https://ui.adsabs.harvard.edu/abs/1974Phy....74..466A}
}

@article{HellerBounds2022ejw,
    author = "Heller, Michal P. and Serantes, Alexandre and Spali{\'n}ski, Micha{\l} and Withers, Benjamin",
    title = "{Rigorous Bounds on Transport from Causality}",
    eprint = "2212.07434",
    archivePrefix = "arXiv",
    primaryClass = "hep-th",
    doi = "10.1103/PhysRevLett.130.261601",
    journal = "Phys. Rev. Lett.",
    volume = "130",
    number = "26",
    pages = "261601",
    year = "2023"
}

@BOOK{peliti_book,
   author = {{Peliti}, L.},
    title = "{Statistical Mechanics in a Nutshell}",
publisher = {Princeton University Press},
   series = {In a nutshell},
     year = {2011}
}

@BOOK{rezzolla_book,
   author = {{Rezzolla}, L. and {Zanotti}, O.},
    title = "{Relativistic Hydrodynamics}",
booktitle = {Relativistic Hydrodynamics, by L.~Rezzolla and O.~Zanotti.~Oxford University Press, 2013.~ISBN-10: 0198528906; ISBN-13: 978-0198528906},
     year = 2013,
    month = sep,
   adsurl = {http://adsabs.harvard.edu/abs/2013rehy.book.....R}
}

@book{Muller_book,
   title =     {Rational Extended Thermodynamics},
   author =    {Muller, I. and Ruggeri, T.},
   publisher = {Springer-Verlag New York},
   year =      {1998},
   edition =   {2nd}
}

@book{MorseFeshbach1953,
  author    = {Morse, Philip M. and Feshbach, Herman},
  title     = {Methods of Theoretical Physics},
  volume    = {1},
  publisher = {McGraw--Hill Book Company},
  address   = {New York},
  year      = {1953},
  note      = {Vol. I}
}

@article{BAGGIOLI20201,
author = {Matteo Baggioli and Mikhail Vasin and Vadim Brazhkin and Kostya Trachenko},
title = {Gapped momentum states},
journal = {Physics Reports},
volume = {865},
pages = {1-44},
year = {2020},
note = {Gapped momentum states},
issn = {0370-1573},
doi = {https://doi.org/10.1016/j.physrep.2020.04.002},
url = {https://www.sciencedirect.com/science/article/pii/S0370157320301198}
}

@book{TeschlBook,
  author    = {Teschl, Gerald},
  title     = {Mathematical Methods in Quantum Mechanics With Applications to Schr\"{o}dinger Operators"},
  series    = {Graduate Studies in Mathematics},
  volume    = {99},
  publisher = {American Mathematical Society},
  address   = {Providence, RI},
  year      = {2009}
}

@article{Fick1855,
  author    = {Adolf Fick},
  title     = {Ueber Diffusion},
  journal   = {Annalen der Physik und Chemie},
  volume    = {94},
  pages     = {59--86},
  year      = {1855},
  doi       = {10.1002/andp.18551700105}
}

@article{NagyOrtizReula1994,
  author       = {Gabriel B. Nagy and Omar E. Ortiz and Oscar A. Reula},
  title        = {The behavior of hyperbolic heat equations’ solutions near their parabolic limits},
  journal      = {Journal of Mathematical Physics},
  volume       = {35},
  number       = {8},
  pages        = {4334--4356},
  year         = {1994},
  doi          = {10.1063/1.530856},
  url          = {https://doi.org/10.1063/1.530856},
}

\label{lastpage}
\end{document}